\documentclass[notitlepage,nofootinbib,preprintnumbers,amssymb,superscriptaddress]{revtex4-1}
\usepackage{amsfonts,amssymb,amsmath,graphicx,color,bm}
\usepackage[linktocpage=true]{hyperref}
\hypersetup{
colorlinks=true,
citecolor=red,
linkcolor=blue,
urlcolor=blue,
}

\def\ear{\end{eqnarray}}
\def\beq{\begin{equation}}             \def\earn{\nonumber \end{eqnarray}}
\def\eeq{\end{equation}}               
\def\bear{\begin{eqnarray}}

\begin{document}

\title{Gauss-Bonnet term corrections in scalar field cosmology}

\author{Igor Fomin}
\email[E-mail:]{ingvor@inbox.ru}
\affiliation{Department of Physics, Bauman Moscow State Technical University,
  Moscow, 105005, Russia}


\begin{abstract}
The influence of non-minimal coupling of a scalar field and the Gauss-Bonnet term on the inflationary stage of evolution of the universe is investigated in this paper. The main cosmological effects of such a coupling were considered.  The deviations between Einstein-Gauss-Bonnet inflation and standard one based on Einstein gravity were determined. The corrections of a weak GB coupling preserving the type of the scalar field potential to standard inflationary models is considered as well.
\end{abstract}

\maketitle

\section{Introduction}

At this stage in the development of theoretical investigations of the early universe, cosmological inflation~\cite{Starobinsky:1980te,Guth:1980zm,Linde:1981mu,Albrecht:1982wi,Einhorn:1980ik,Sato:1981ds,Sato:1981ds,
Linde:1983gd} seems to be the most convincing theory.
The first models of cosmological inflation were based principally on Einstein gravity and the assumption that there is some scalar field $\phi$ as an ideal barotropic fluid with negative pressure at the inflationary stage of the evolution of early universe~\cite{Guth:1980zm,Linde:1981mu,Albrecht:1982wi,Einhorn:1980ik,Sato:1981ds,Sato:1981ds,
Linde:1983gd}.
Also, according to the theory of cosmological perturbations, quantum fluctuations of a scalar field induce corresponding perturbations of the metric, which give rise to a large-scale structure of the universe and relic gravitational waves \cite{Mukhanov:1990me}.
At the moment, a large number of different models of cosmological inflation with canonical scalar fields based on Einstein gravity are considered to describe the inflationary stage of the evolution of universe \cite{Martin:2013tda}.

Another possibility for constructing the cosmological models of early universe is using modified gravity theories which include the higher-order curvature terms~\cite{Starobinsky:1980te,Nojiri:2010wj,Clifton:2011jh,Baumann:2014nda,Nojiri:2017ncd,Ishak:2018his} which can be associated with quantum effects in the low-energy limit of string theory and supergravity.
One such a correction is the Gauss-Bonnet term (scalar) $R^{2}_{\rm GB}= R_{\mu\nu\rho\sigma} R^{\mu\nu\rho\sigma} - 4 R_{\mu\nu}R^{\mu\nu} + R^2$ which arises in the low-energy effective action for the heterotic strings~\cite{Zwiebach:1985uq,Zumino:1985dp,Boulware:1985wk,Boulware:1986dr,Gasperini:1996fu}, and also appears in the second order of Lovelock gravity theory \cite{Lovelock:1971yv}.

Cosmological models with the Gauss-Bonnet (GB) term  in four-dimensional Friedmann universe was considered earlier in a large number of works (for example, see~\cite{Kanti:1998jd,Nojiri:2005vv,Nojiri:2005jg,Jiang:2013gza,Kanti:2015pda,Hikmawan:2015rze,vandeBruck:2015gjd,
Mathew:2016anx,Granda:2017oku,vandeBruck:2017voa,Heydari-Fard:2016nlj,Sberna:2017xqv,Yi:2018gse,Chakraborty:2018scm,
Pozdeeva:2019agu}).
The important property of such a models is that the GB-term affects the cosmological dynamics in four-dimensional space-time only for the case of non-minimal coupling of this term with a scalar field ~\cite{Kanti:1998jd,Nojiri:2005vv,Nojiri:2005jg,Jiang:2013gza,Kanti:2015pda,Hikmawan:2015rze,vandeBruck:2015gjd,
Mathew:2016anx,Granda:2017oku,vandeBruck:2017voa,Heydari-Fard:2016nlj,Sberna:2017xqv,Yi:2018gse,Pozdeeva:2019agu}
that can be defined by some coupling function $\xi(\phi)$.

The evolution of cosmological perturbations and their corresponding parameters for Einstein-Gauss-Bonnet (EGB) inflationary models were considered in papers
\cite{Satoh:2008ck,Guo:2009uk,Guo:2010jr,DeFelice:2011uc,DeFelice:2011jm,Koh:2014bka,Koh:2016abf,
Bhattacharjee:2016ohe,Wu:2017joj,Odintsov:2018zhw}.
In this case, it should be noted that the non-minimal coupling of a scalar field and the Gauss-Bonnet scalar allows to verify cosmological inflationary models from observational constraints on the values of cosmological perturbation parameters \cite{Ade:2015xua,Akrami:2018odb}, in contrast to some models constructed based on Einstein gravity only
due to difference in the evolution of perturbations at the inflationary stage \cite{Mathew:2016anx,Guo:2009uk,Guo:2010jr,Koh:2014bka,Koh:2016abf}.

An important difference between inflationary models based on Einstein-Gauss-Bonnet gravity and ones based on GR is the dependence of the velocities of the propagation of cosmological perturbations on cosmic time \cite{Satoh:2008ck,Guo:2009uk,Guo:2010jr,DeFelice:2011uc,DeFelice:2011jm,Koh:2014bka,Koh:2016abf,
Bhattacharjee:2016ohe,Wu:2017joj,Odintsov:2018zhw}, that implies the deviations of these velocities from the speed of light in a vacuum for EGB-inflation.

A common method for analyzing cosmological models based on modified gravity theories is conformal transformations of a space-time metric that bring the initial action to the Einstein-Hilbert form with corresponding transformations of the material components \cite{Nojiri:2010wj,Clifton:2011jh,Baumann:2014nda,Nojiri:2017ncd}. The explicit form of the transformations allows one to compare models based on General Relativity and its modifications. However, for the case of Einstein-Gauss-Bonnet gravity, no such a transformations were found.

In papers~\cite{Fomin:2017sqz,Fomin:2017vae,Fomin:2017qta,Fomin:2018typ,Fomin:2019yls} it was proposed to consider the relationship between standard inflationary models and EGB-inflation directly from the equations of cosmological dynamics in flat four-dimensional Friedmann-Robertson-Walker (FRW) space-time, which is sufficient for comparing such a models, since this type of geometry is the basis for constructing phenomenologically correct cosmological of models \cite{Ade:2015xua,Akrami:2018odb,Ehlers:1966ad,Clarkson:2003ts}. Also, this approach was used to analyze cosmological inflationary models based on the other modifications of Einstein gravity \cite{Fomin:2019yls,Fomin:2018blx,Fomin:2017sbt}.
Thus, the application of the approach based on a connection between Einstein-Gauss-Bonnet gravity and General Relativity in relevant cosmological models makes it possible to evaluate the effects of the non-minimal coupling of a scalar field and the Gauss-Bonnet term.

The aim of this work is to develop the method of analysis of Gauss-Bonnet term corrections to standard inflationary models which was proposed in~\cite{Fomin:2017sqz,Fomin:2017vae,Fomin:2017qta,Fomin:2018typ,Fomin:2019yls}.
The article is organized as follows. In Sec. \ref{section2}, the difference between equations of cosmological dynamics for the case of Einstein-Gauss-Bonnet gravity and General Relativity in flat four-dimensional FRW space-time is considered.
In Sec. \ref{section3}, this difference is defined in terms of the deviation parameters, and estimates of the influence of the non-minimal coupling of a scalar field and the Gauss-Bonnet term on the main parameters of the background cosmological dynamics are given. It was further obtained that the slow-roll conditions imply a weak effect of such a coupling on cosmological dynamics.
This result was applied in Sec. \ref{section4} for the analysis of EGB-inflationary models with a weak coupling and the parameterization of the Gauss-Bonnet term corrections through a coupling constant.
In Sec. \ref{section5}, different cosmological inflationary models with weak a GB-coupling were considered, and the main effects of such a coupling on the inflationary parameters were determined. In conclusion, the results of this work were discussed.

\section{The cosmological models based on the Einstein-Gauss-Bonnet gravity}\label{section2}

The models of cosmological inflation with Einstein gravity can be considered on the basis of the action
\cite{Guth:1980zm,Linde:1981mu,Albrecht:1982wi,Einhorn:1980ik,Sato:1981ds,Sato:1981ds,
Linde:1983gd}
\begin{eqnarray}\label{S-1}
S_{E} = \int_{\mathcal{M}} d^4x\sqrt{-g}\Big[\frac{1}{2} R -
 \frac{1}{2}g^{\mu\nu}\partial_{\mu}\phi_{E} \partial_{\nu} \phi_{E}- V_{E}(\phi_{E})\Big],
\label{actionE}
\end{eqnarray}
and for inflationary models with additional non-minimal coupling of a scalar field and the Gauss-Bonnet term, the action is
\cite{Kanti:1998jd,Nojiri:2005vv,Nojiri:2005jg,Jiang:2013gza,Kanti:2015pda,Hikmawan:2015rze,vandeBruck:2015gjd,
Mathew:2016anx,Granda:2017oku,vandeBruck:2017voa,Heydari-Fard:2016nlj,Sberna:2017xqv,Yi:2018gse,Pozdeeva:2019agu}
\begin{eqnarray}
S_{GB} = \int_{\mathcal{M}} d^4x\sqrt{-g}\Big[\frac{1}{2} R -
 \frac{1}{2}g^{\mu\nu}\partial_{\mu}\phi_{GB} \partial_{\nu} \phi_{GB}
 - V_{GB}(\phi_{GB})-\frac12\xi(\phi_{GB}) R_{\rm GB}^2\Big],
\label{actionGB}
\end{eqnarray}
where $\phi$ is a scalar field with the potential $V(\phi)$, $R$ the Ricci scalar curvature of the space-time $\mathcal{M}$, $R^{2}_{\rm GB}= R_{\mu\nu\rho\sigma} R^{\mu\nu\rho\sigma} - 4 R_{\mu\nu}R^{\mu\nu} + R^2$  the Gauss-Bonnet  term and $\xi(\phi_{GB})$ is a coupling function. Index ``$E$'' denotes Einstein gravity, and index ``$GB$'' means Einstein-Gauss-Bonnet gravity.

The background dynamic equations corresponding to the action (\ref{actionGB}) in a spatially flat four-dimensional Friedmann-Robertson-Walker space-time
\begin{equation}
ds^2=-dt^2+a^{2}(t)\left(dx^{2}+dy^{2}+dz^{2}\right),
\end{equation}
in the system of units $8\pi G=c=1$, are \cite{Kanti:1998jd,Nojiri:2005vv,Nojiri:2005jg,Jiang:2013gza,Kanti:2015pda,Hikmawan:2015rze,vandeBruck:2015gjd,
Mathew:2016anx,Granda:2017oku,vandeBruck:2017voa,Heydari-Fard:2016nlj,Sberna:2017xqv,Yi:2018gse,Pozdeeva:2019agu}
\begin{eqnarray}
\label{beq1a}
&&3H^{2}_{GB}=\frac{1}{2}\dot{\phi}^{2}_{GB}+V_{GB}+12\dot{\xi}H^{3}_{GB},\\
\label{beq2b}
&&\dot{\phi}^{2}_{GB}=-2\dot{H}_{GB}+4\ddot{\xi}H^{2}_{GB}+4\dot{\xi}H_{GB}(2\dot{H}_{GB}-H^{2}_{GB}),\\
\label{beq3}
&&\ddot{\phi}_{GB} + 3 H_{GB} \dot{\phi}_{GB} + \frac{\partial V_{GB}(\phi_{GB})}{\partial\phi_{GB}} +
12H^2_{GB} \left(\dot{H}_{GB}+H^2_{GB}\right)\frac{\partial\xi(\phi_{GB})}{\partial\phi_{GB}}= 0,
\end{eqnarray}
where a dot represents a derivative with respect to the cosmic time $t$, $H \equiv \dot{a}/a$ denotes the Hubble parameter and
$a=a(t)$ is a scale factor.

Since the equation (\ref{beq3}) can be derived from (\ref{beq1a})--(\ref{beq2b}), one can consider the dynamic equations (\ref{beq1a})--(\ref{beq3}) in the following form
\begin{eqnarray}
 \label{EQG1}
&&V_{GB}(\phi_{GB})=3H^{2}_{GB}+\dot{H}_{GB}-10H^{3}_{GB}\dot{\xi}-2H^{2}_{GB}\ddot{\xi}
-4H_{GB}\dot{H}_{GB}\dot{\xi},\\
\label{EQG2}
&&\frac{1}{2}\dot{\phi}^{2}_{GB}=-\dot{H}_{GB}-2H^{3}_{GB}\dot{\xi}+
4H_{GB}\dot{H}_{GB}\dot{\xi}+2H^{2}_{GB}\ddot{\xi}.
\end{eqnarray}

If $\xi$ is a constant, then equations (\ref{EQG1})--(\ref{EQG2}) are reduced to those for standard inflationary dynamic equations corresponding to the minimal coupling or the case of Einstein gravity
\begin{eqnarray}
 \label{E1}
&&V_{E}(\phi_{E})=3H^{2}_{E}+\dot{H}_{E},\\
\label{E2}
&&\frac{1}{2}\dot{\phi}^{2}_{E}=-\dot{H}_{E},
\end{eqnarray}
which can be obtained from the action (\ref{S-1}).

Thus, the non-minimal coupling of a scalar field with the Gauss-Bonnet scalar changes the evolution of a field itself, its potential and the dynamics of the universe's expansion. The connection between EGB-inflation and standard one in a spatially flat four-dimensional Friedmann-Robertson-Walker space-time can be determined from the equations (\ref{EQG1})--(\ref{EQG2}) and (\ref{E1})--(\ref{E2}). Therefore, the first step in this analysis is to explicitly identify such a connection.

\section{The deviations between Einstein-Gauss-Bonnet inflation and standard one}\label{section3}

To obtain the explicit form of the connection between background inflationary parameters of standard inflation $\{\phi_{E},V_{E},H_{E}\}$ and Einstein-Gauss-Bonnet inflation $\{\phi_{GB},V_{GB},H_{GB}\}$  one can use the relation between the Hubble parameters $H_{E}$ and $H_{GB}$, which was considered in the papers~\cite{Fomin:2017sqz,Fomin:2017vae,Fomin:2017qta,Fomin:2018typ,Fomin:2019yls}, namely
\begin{equation}
\label{connection}
H_{E}=H_{GB}\left(1-2\dot{\xi}H_{GB}\right).
\end{equation}

Thus, using this connection, one can rewrite the equations  (\ref{EQG1})--(\ref{EQG2}) in the following form
\begin{eqnarray}
\label{ex3}
&&V_{GB}(\phi_{GB})=-2H^{2}_{GB}+5H_{E}H_{GB}+\dot{H}_{E},\\
\label{ex4}
&&\frac{1}{2}\dot{\phi}^{2}_{GB}=-\dot{H}_{E}+H_{E}H_{GB}-H^{2}_{GB}.
\end{eqnarray}

As one can see, on the one hand, the substituting of expression (\ref{connection}) into equations (\ref{ex3})--(\ref{ex4}) leads to equations (\ref{EQG1})--(\ref{EQG2}) with $\phi_{GB}\neq\phi_{E}$ and $V_{GB}\neq V_{E}$ for $\xi\neq const$, on the other hand, for the case $\xi=const$, the expression (\ref{connection}) leads to $H_{GB}=H_{E}$, and dynamic equations
(\ref{ex3})--(\ref{ex4}) are reduced to ones (\ref{E1})--(\ref{E2}) for inflation based of Einstein gravity, that implies
$\phi_{GB}=\phi_{E}$ and $V_{GB}=V_{E}$.

Therefore, the GB-term corrections to standard inflation can be defined as follows
\begin{eqnarray}
\label{COR1}
&&V_{GB}=V_{E}+\Delta_{V},\\
\label{COR2}
&&X_{GB}=X_{E}+\Delta_{X},\\
\label{COR3}
&&H_{GB}=H_{E}+\Delta_{H},
\end{eqnarray}
where $X_{GB}\equiv\frac{1}{2}\dot{\phi}^{2}_{GB}$ and $X_{E}\equiv\frac{1}{2}\dot{\phi}^{2}_{E}$ are the kinetic energies of a scalar field in EGB-inflation and standard one.

The functions $\Delta_{V}$, $\Delta_{X}$ and $\Delta_{H}$ characterize the GB-term corrections to background inflationary parameters, and, therefore, they can be considered as the deviation parameters between EGB-inflation and standard one based on GR. The explicit expressions for deviation parameters one can obtain from equations (\ref{EQG1})--(\ref{E2}) in the following form
\begin{eqnarray}
\label{DP1}
&&\Delta_{V}=-2H^{2}_{GB}+5H_{E}H_{GB}-3H^{2}_{E}=2H^{3}_{GB}\dot{\xi}(1-6H_{GB}\dot{\xi}),\\
\label{DP2}
&&\Delta_{X}=-H_{GB}(H_{GB}-H_{E})=-2H^{3}_{GB}\dot{\xi},\\
\label{DP3}
&&\Delta_{H}=H_{GB}-H_{E}=2H^{2}_{GB}\dot{\xi},
\end{eqnarray}

These parameters are connected by the relations
\begin{equation}
\label{REL}
\Delta_{V}=-3\Delta^{2}_{H}-\Delta_{X},~~~\Delta_{X}=-H_{GB}\Delta_{H}.
\end{equation}

Since the deviation parameters (\ref{DP1})--(\ref{DP3}) can be both positive and negative, the character of the influence of non-minimal GB coupling depends on the choice of a specific model of cosmological inflation.

\subsection{The influence of GB-term on a scalar field}

As the first application of proposed approach, one can determine the change in the characteristics of a scalar field inspired by non-minimal GB coupling.

The influence of such a coupling on the pressure of a scalar field $p=X-V$ can be found as the difference between
pressures for EGB-inflation and standard one
\begin{equation}
\label{PRESSURE}
p_{GB}-p_{E}=X_{GB}-X_{E}-V_{GB}+V_{E}=\Delta_{X}-\Delta_{V}=3\Delta^{2}_{H}+2\Delta_{X}.
\end{equation}
As one can see, this difference depends on the model's type of inflation.

However, for the difference between energy densities of a scalar field $\rho=X+V$ one has
\begin{equation}
\label{ENERGY}
\rho_{GB}-\rho_{E}=X_{GB}-X_{E}+V_{GB}-V_{E}=\Delta_{X}+\Delta_{V}=-3\Delta^{2}_{H}<0,
\end{equation}
for any inflationary model.

Therefore, in four-dimensional spatially flat Friedmann-Robertson-Walker space-time the non-minimal coupling of a scalar field with the Gauss-Bonnet term leads to decrease of it's energy density.

Further, one can write a state parameter of a coupled scalar field as
\begin{equation}
\label{STATEPARAMETERGB}
w_{GB}=\frac{p_{GB}}{\rho_{GB}}=w_{E}+\frac{3\Delta^{2}_{H}(p_{E}+\rho_{E})+2\Delta_{X}\rho_{E}}
{(\rho_{E}-3\Delta^{2}_{H})\rho_{E}}
=w_{E}+\frac{2}{3}\left(\frac{\Delta^{2}_{H}\epsilon_{E}+\Delta_{X}}
{H^{2}_{E}-\Delta^{2}_{H}}\right)=-1-\frac{2}{3}\left(\frac{\dot{H}_{E}-\Delta_{X}}{H^{2}_{E}-\Delta^{2}_{H}}\right),
\end{equation}
where
\begin{equation}
\label{STATEPARAMETER}
w_{E}=\frac{p_{E}}{\rho_{E}}=-1-\frac{2\dot{H}_{E}}{3H^{2}_{E}}=-1+\frac{2}{3}\epsilon_{E},
\end{equation}
is a state parameter and $\epsilon_{E}=-\dot{H}_{E}/H^{2}_{E}$ is slow-roll parameter for the case of Einstein gravity.

Thus, in general case, the  non-minimal coupling of a scalar field with the Gauss-Bonnet scalar can significantly change the equation of state of a scalar field. Nevertheless, when evaluating such a changes, the condition $\rho_{GB}>0$ should be taken into account, which will be considered further. Also, for the case $\dot{\xi}=0$ one has $\Delta_{X}=\Delta_{H}=0$ and the state parameter $w_{GB}$ is reduced to $w_{E}$.

\subsection{The influence of GB-term on the background dynamics}

The influence of such a coupling on the dynamics can be qualitatively estimated by the sign of $\dot {\xi}$. In the case of decreasing coupling function $\xi (t)$ $(\dot{\xi}<0)$ one has $H_{GB}-H_ {E}<0$, that means a decrease in the rate of expansion of the universe relative to the standard inflationary models and one has the inverse effect (acceleration) in the case of the growth of coupling function $\xi (t)$ $(\dot{\xi}>0)$.

This effect can be also quantified by the difference in the $e$-folds numbers which changes as
\begin{equation}
N_{GB}-N_{E}=\int^{t_{e}}_{t_{i}}(H_{GB}-H_{E})dt=\int^{t_{e}}_{t_{i}}\Delta_{H}dt,
\end{equation}
where $t_{i}$ and $t_{e}$ are the times of the beginning and the end of inflationary stage.

From the conditions of positive energy density of a scalar field $\rho_{GB}>0$, expansion of the universe $H_{E}>0$, $H_{GB}>0$ and expression $\rho_{E}=X_{E}+V_{E}=3H^{2}_{E}$ one has following restriction on the Hubble parameter for EGB-inflation
\begin{equation}
\label{RESTRICTIONH}
\rho_{GB}=\rho_{E}-3\Delta^{2}_{H}>0,~~~~~0<H_{GB}<2H_{E},
\end{equation}
and the restrictions on the deviation parameters
\begin{equation}
\label{RESTRICTIONH1}
-H_{E}<\Delta_{H}<H_{E},~~~~~-2H^{2}_{E}<\Delta_{X}<2H^{2}_{E},
\end{equation}
which limits changes in the state parameter of a scalar field (\ref{STATEPARAMETERGB}) as well.

In terms of the slow-roll parameter $\epsilon$ which characterize the dynamics of the inflationary stage, this restriction can be formulated as
\begin{equation}
\label{RESTRICTIONE}
\frac{1}{2}\epsilon_{E}<\epsilon_{GB}<1,
\end{equation}
since $\epsilon=1$ is the condition for completing the stage of cosmological inflation.

Also, from (\ref{RESTRICTIONH}) one has the following restriction on the increment of $e$-folds number
\begin{equation}
N_{GB}-N_{E}<N_{E},~~~~~N_{GB}<2N_{E},
\end{equation}
for Einstein-Gauss-Bonnet inflation compared to standard one.

Thus, the results obtained mean that the non-minimal coupling of a scalar field and the Gauss-Bonnet scalar can accelerate the rate of expansion of the Friedmann universe by less than two times.

\subsection{The influence of GB-term on the parameters of cosmological perturbations}\label{GBPERT}

In accordance with the theory of cosmological perturbations, quantum fluctuations of the scalar field generate the corresponding perturbations of the space-time metric during the inflationary stage. In the linear order of cosmological perturbation theory, the observed anisotropy and polarization of cosmic microwave background radiation (CMB) \cite{Ade:2015xua,Akrami:2018odb} are explained by the influence of two types of perturbations, namely, scalar and tensor ones. The third type of perturbations (vector perturbations) quickly decay in the process of accelerated expansion of the early universe \cite{Mukhanov:1990me}.

The calculation of cosmological perturbation parameters for inflationary models with taking into account the non-minimal
coupling of a scalar field and the Gauss-Bonnet scalar was carried out in many works \cite{Guo:2009uk,Guo:2010jr,Koh:2014bka,Koh:2016abf,Bhattacharjee:2016ohe,Wu:2017joj,
Odintsov:2018zhw,Satoh:2008ck,DeFelice:2011uc,DeFelice:2011jm} before.

To compare the predictions of the inflationary model with the observational data of CMB anisotropy, it suffices to consider two parameters of cosmological perturbations, namely, the spectral index of scalar perturbations $n_{S}$ and the tensor-scalar ratio $r$ which is the ratio of the squared amplitudes of tensor and scalar perturbations \cite{Guo:2009uk,Guo:2010jr,Koh:2014bka,Koh:2016abf,Bhattacharjee:2016ohe,Wu:2017joj,
Odintsov:2018zhw,Satoh:2008ck,DeFelice:2011uc,DeFelice:2011jm}.

The expressions for these parameters on the crossing of the Hubble radius ($k=aH$) can be written as follows \cite{Guo:2009uk,Guo:2010jr,Koh:2014bka,Koh:2016abf,Bhattacharjee:2016ohe,Wu:2017joj,
Odintsov:2018zhw,Satoh:2008ck,DeFelice:2011uc,DeFelice:2011jm}
\begin{eqnarray}
\label{NSGB}
&&n_{S(GB)}-1=-2\epsilon_{GB}-\frac{2\epsilon_{GB}(2\epsilon_{GB}-2\delta_{GB})+\Delta_{2}}
{2\epsilon_{GB}-\Delta_{1}},\\
\label{RGB}
&&r_{GB}=8\left(2\epsilon_{GB}-\Delta_{1}\right),
\end{eqnarray}
where the slow-roll  parameters and deviation ones for EGB-inflation are defined as
\begin{eqnarray}
\label{EPSDELGB}
&&\epsilon_{GB}=-\frac{\dot{H}_{GB}}{H^{2}_{GB}},~~~
\delta_{GB}=-\frac{\ddot{H}_{GB}}{2\dot{H}_{GB}H_{GB}}=
\epsilon_{GB}-\frac{\dot{\epsilon}_{GB}}{2\epsilon_{GB}H_{GB}},\\
\label{DELTA}
&&\Delta_{1}=4\dot{\xi}H_{GB},~~~\Delta_{2}=-4\ddot{\xi}+\Delta_{1}\epsilon_{GB}.
\end{eqnarray}

On the basis of expression (\ref{connection}), one can redefine the deviation parameters as
\begin{eqnarray}
\label{DELTA1}
&&\Delta_{1}=2\left(1-\frac{H_{E}}{H_{GB}}\right)=2\left(1-\frac{H_{E}}{H_{E}+\Delta_{H}}\right),\\
\label{DELTA2}
&&\Delta_{2}=2\left(\frac{\dot{H}_{E}}{H^{2}_{GB}}-\frac{\dot{H}_{GB}H_{E}}{H^{3}_{GB}}\right)=
-\frac{\dot{\Delta}_{1}}{H_{GB}}.
\end{eqnarray}

The deviation parameters $\Delta_{1}$ and $\Delta_{2}$ are connected with slow-roll parameters $\delta_{1}$ and $\delta_{2}$ which are usually used to define the GB-term corrections \cite{Koh:2014bka,Koh:2016abf}
as follows\footnote{In papers \cite{Guo:2009uk,Guo:2010jr,Wu:2017joj} the other parameter $\tilde{\delta}_{2}=\delta_{2}-\epsilon_{GB}$ corresponding to the hierarchy $\tilde{\delta}_{i+1}=d\ln|\tilde{\delta}_{i}|/d\ln a$ ($i\geq1$) was used.}
\begin{eqnarray}
\label{DELTA1P}
&&\Delta_{1}=\delta_{1},\\
\label{DELTA2R}
&&\Delta_{2}=-\delta_{1}\delta_{2}+\delta_{1}\epsilon_{GB},\\
\label{DELTA3R}
&&\delta_{2}=\frac{\ddot{\xi}}{\dot{\xi} H_{GB}}=\epsilon_{GB}-\frac{\Delta_{2}}{\Delta_{1}}.
\end{eqnarray}

The parameters (\ref{DELTA1}) and (\ref{DELTA2}) are considered instead of $\delta_{1}$ and $\delta_{2}$,  in framework of proposed approach, for the convenience of analyzing the deviations between EGB-inflation and the standard inflationary scenarios, since for $\xi=const$, parameters $\delta_{2}$ and $\tilde{\delta}_{2}$ have an undefined values.

From equations (\ref{NSGB})--(\ref{DELTA2}) one has following expressions for the spectral index of scalar perturbations and tensor-to-scalar ratio in the case of EGB-inflation
\begin{eqnarray}
\label{NSGB2}
&&n_{S(GB)}-1=-2\epsilon_{GB}-\frac{2\dot{\epsilon}_{GB}-\dot{\Delta}_{1}}{(2\epsilon_{GB}-\Delta_{1})H_{GB}}=
-(\Delta_{1}+\Delta_{3})-\frac{\dot{\Delta}_{3}}{\Delta_{3}H_{GB}},\\
\label{RGB2}
&&r_{GB}=8\left(2\epsilon_{GB}-\Delta_{1}\right)=8\Delta_{3},
\end{eqnarray}
where
\begin{equation}
\label{DELTA3}
\Delta_{3}=2\epsilon_{GB}-\Delta_{1}.
\end{equation}

Also, one can define the connection between background deviation parameters $\{\Delta_{H},\Delta_{X},\Delta_{V}\}$ and ones corresponding to the Gauss-Bonnet term corrections to the parameters of cosmological perturbations $\{\Delta_{1},\Delta_{2},\Delta_{3}\}$.

Firstly, from equations (\ref{DP1})--(\ref{DP3}) one can obtain
\begin{eqnarray}
\label{DPH1}
&&H_{GB}=-\frac{\Delta_{X}}{\Delta_{H}},\\
\label{DPH2}
&&H_{E}=-\frac{\Delta_{X}+\Delta^{2}_{H}}{\Delta_{H}}.
\end{eqnarray}

Further, from expressions (\ref{DELTA1}), (\ref{DELTA2}) and (\ref{DELTA3}), taking into account (\ref{DPH1})--(\ref{DPH2}), one has
\begin{eqnarray}
\label{DPCP1}
&&\Delta_{1}=-2\frac{\Delta^{2}_{H}}{\Delta_{X}},\\
\label{DPCP2}
&&\Delta_{2}=-2\frac{\Delta_{H}}{\Delta_{X}}\frac{d}{dt}\left(\frac{\Delta^{2}_{H}}{\Delta_{X}}\right),\\
\label{DPCP3}
&&\Delta_{3}=2\frac{\Delta^{2}_{H}}{\Delta_{X}}-2\frac{d}{dt}\left(\frac{\Delta_{H}}{\Delta_{X}}\right).
\end{eqnarray}

Also, it as possible to write the relations (\ref{DPCP1})--(\ref{DPCP3}) in terms of third background deviation parameter $\Delta_{V}$ on the basis of the connections (\ref{REL}).
Thus, the relations  (\ref{DPH1})--(\ref{DPCP3}) allow one to determine the parameters $\{\Delta_{1},\Delta_{2},\Delta_{3}\}$ from the background deviation parameters $\{\Delta_{H},\Delta_{X},\Delta_{V}\}$
that will be used in following analysis of inflationary models.

The expressions (\ref{NSGB})--(\ref{RGB}) were obtained in \cite{Guo:2009uk,Guo:2010jr,Koh:2014bka,Koh:2016abf,Bhattacharjee:2016ohe,Wu:2017joj,
Odintsov:2018zhw,Satoh:2008ck,DeFelice:2011uc,DeFelice:2011jm} taking into account the slow-roll conditions, which can be written as
\begin{eqnarray}
\label{SLOWROLLGB1}
&&\epsilon_{GB}\ll1,~~~~\delta_{GB}\ll1,\\
\label{SLOWROLLGB}
&&\Delta_{1}\ll1,~~~~\Delta_{2}\ll1,~~~~\Delta_{3}\ll1.
\end{eqnarray}

For the case of Einstein gravity $\dot{\xi}=0$ one has $H_{GB}=H_{E}$, $\epsilon_{GB}=\epsilon_{E}$,
$\delta_{GB}=\delta_{E}$ and $\Delta_{1}=\Delta_{2}=0$, and expressions for the parameters of cosmological perturbations are reduced to
\begin{eqnarray}
\label{NSE}
&&n_{S(E)}-1=-4\epsilon_{E}+2\delta_{E},\\
\label{RE}
&&r_{E}=16\epsilon_{E},
\end{eqnarray}
that corresponds to the result obtained for the standard inflation \cite{Liddle:1994dx,Lyth:1998xn,Chervon:2008zz} with the slow-roll conditions
\begin{equation}
\label{SLOWROLLE}
\epsilon_{E}\ll1,~~~~\delta_{E}\ll1,
\end{equation}
where the slow-roll parameters for the standard inflation based on Einstein gravity are
\begin{equation}
\label{EPSDELE}
\epsilon_{E}=-\frac{\dot{H}_{E}}{H^{2}_{E}},~~~~~
\delta_{E}=-\frac{\ddot{H}_{E}}{2\dot{H}_{E}H_{E}}=
\epsilon_{E}-\frac{\dot{\epsilon}_{E}}{2\epsilon_{E}H_{E}}.
\end{equation}

The parameters of cosmological perturbations must satisfy the following observational constraints \cite{Ade:2015xua,Akrami:2018odb}
\begin{eqnarray}
\label{PLANCK1}
&&n_{S}=0.9663\pm 0.0041,\\
\label{PLANCK2}
&&r<0.065,
\end{eqnarray}
that defines the method of the verification of cosmological inflationary models.

From restrictions (\ref{RESTRICTIONH})--(\ref{RESTRICTIONH1}) and definition (\ref{DELTA1}) one has
the condition $\Delta_{1}<1$. However, the equation (\ref{RGB2}) leads to $\Delta_{1}<2\epsilon_{GB}\ll1$. Thus, the expressions (\ref{NSGB})--(\ref{RGB}) are valid only for the case of satisfying conditions  (\ref{SLOWROLLGB}).
A consequence of the conditions (\ref{SLOWROLLGB}), taking into account the expressions (\ref{DELTA1})--(\ref{DELTA2}), is the weak influence of non-minimal coupling of a scalar field and the Gauss-Bonnet scalar on cosmological dynamics $\Delta_{H}\ll H_{E}$.

Thus, the slow-roll conditions for EGB-inflation correspond to the interpretation of non-minimal coupling of a scalar field and the Gauss-Bonnet scalar as a small quantum corrections to the main dynamical effects determined by Einstein gravity at the inflationary stage of the evolution of early universe, which can be called as a weak GB coupling.

\subsection{The influence of GB-term on the velocities of cosmological perturbations}\label{GBPERT}

To analyze the influence of non-minimal coupling on the velocities of cosmological perturbations one can use their exact expressions which were considered, for example,  in \cite{Koh:2014bka,Koh:2016abf}.

The velocity of the scalar perturbations can be expressed as follows
\begin{equation}
\label{SKVEL}
c^{2}_{S}=1-\frac{[4\epsilon_{GB}-\delta_{1}(1-4\epsilon_{GB}-\delta_{2})]\Delta^{2}}
{4\epsilon_{GB}-2\delta_{1}-2\delta_{1}(2\epsilon_{GB}-\delta_{2})+3\delta_{1}\Delta},
\end{equation}
where $\Delta=\frac{\delta_{1}}{1-\delta_{1}}$.

In terms of the deviation parameters $\Delta_{1}$ and $\Delta_{2}$ the expression (\ref{SKVEL}) can be noted as
\begin{equation}
\label{SKVELD}
c^{2}_{S}=\frac{7\Delta^{3}_{1}\epsilon_{GB}+4\Delta^{3}_{1}+\Delta^{2}_{1}\Delta_{2}-12\Delta^{2}_{1}
-7\Delta^{2}_{1}-4\Delta_{1}\Delta_{2}+10\Delta_{1}\epsilon_{GB}+2\Delta_{1}+2\Delta_{2}-4\epsilon_{GB}}
{(2\Delta^{2}_{1}\epsilon_{GB}+5\Delta^{2}_{1}+2\Delta_{1}\Delta_{2}-6\Delta_{1}\epsilon_{GB}-
2\Delta_{1}-2\Delta_{2}+4\epsilon_{GB})(\Delta_{1}-1)}.
\end{equation}

On the basis of the slow-roll conditions (\ref{SLOWROLLGB1})--(\ref{SLOWROLLGB}) for a weak GB coupling, after neglecting the small terms second and higher orders, from expression (\ref{SKVELD}) one has $c^{2}_{S}\simeq 1$.

The expression for the velocity of tensor perturbations (relic gravitational waves) for the case of EGB-gravity is \cite{Koh:2014bka,Koh:2016abf}
\begin{equation}
\label{WGVEL}
c^{2}_{g}=1+\frac{\delta_{1}(1-\delta_{2})}{1-\delta_{1}}=
1+\frac{\Delta_{1}-\Delta_{1}\epsilon_{GB}+\Delta_{2}}{1-\Delta_{1}}.
\end{equation}

After neglecting the second order small term $\Delta_{1}\epsilon_{GB}$, from expression (\ref{WGVEL}) one has
\begin{equation}
\label{WGVELD}
c^{2}_{g}\simeq\frac{1+\Delta_{2}}{1-\Delta_{1}}.
\end{equation}

Thus, the parameters $\Delta_{1}$ and $\Delta_{2}$ define the small deviations of  the velocity of tensor perturbations from the speed of light in vacuum $c$ (in chosen system of units $c=1$). It should be noted, that the velocity of gravitational waves $c_{g}=1$ corresponds to a limited class of gravity theories, including General Relativity \cite{Ezquiaga:2017ekz}.

Based on detection of gravitational waves from neutron star merging GW$170817$ event \cite{GBM:2017lvd} in modern era of the universe's evolution one has the following restriction on the value of their velocity $|c_{g}-1|\leq5\times10^{-16}$.
In papers \cite{Odintsov:2020zkl,Odintsov:2020sqy,Odintsov:2019clh} this result was extrapolated to the relic gravitational waves on inflationary stage, and the conditions on the parameters $\delta_{1}$, $\delta_{2}$ and corresponding coupling function $\xi$ from expression (\ref{WGVELD}) were found as well.
Another approach to the analysis of EGB cosmological inflationary models, in which the deviations with General Relativity quickly decreased during the expansion of the early universe, which gives a correspondence with $c_{g}=1$  was considered in \cite{Fomin:2019yls}.

In this paper, we will consider a special choice of the deviation parameter $\Delta_{H}$ that implies no difference between EGB gravity and General Relativity for the case of a pure exponential expansion of the universe. Such a choice leads to a description of the second accelerated expansion of the universe in modern era based on Einstein gravity.

\section{Inflation with a weak coupling of a scalar field and Gauss-Bonnet scalar}\label{section4}

The weak influence of non-minimal coupling of the Gauss-Bonnet scalar and a scalar field
on cosmological dynamics can be defined in terms of the deviation parameter $\Delta_{H}$ as
\begin{equation}
\label{WEAKC}
H_{GB}=H_{E}+\Delta_{H}, ~~~\Delta_{H}\ll H_{E}.
\end{equation}

After substituting expression (\ref{WEAKC}) into (\ref{DP2}) and (\ref{REL})
with neglecting the second order terms ${\mathcal O}(\Delta^{2}_{H})$ one has
\begin{eqnarray}
\label{DELTX}
&&\Delta_{X}=-H_{E}\Delta_{H}-\Delta^{2}_{H}\simeq-H_{E}\Delta_{H},\\
&&\Delta_{V}=H_{E}\Delta_{H}-2\Delta^{2}_{H}\simeq H_{E}\Delta_{H}.
\end{eqnarray}

Thus, one can obtain the expressions of the energy density and pressure of a scalar field for a weak GB coupling from (\ref{PRESSURE})--(\ref{ENERGY}) in the following form
\begin{eqnarray}
&&\rho_{GB}=\rho_{E}+\Delta_{X}+\Delta_{V}\simeq\rho_{E}=3H^{2}_{E},\\
&&p_{GB}=p_{E}+2\Delta_{X}+3\Delta^{2}_{H}\simeq p_{E}+2\Delta_{X}\simeq
-3H^{2}_{E}-2\dot{H}_{E}+2H_{E}\Delta_{H}.
\end{eqnarray}

The state parameter (\ref{STATEPARAMETER}) for a weak GB coupling is
\begin{equation}
w_{GB}\simeq-1+\frac{2}{3}\left(\epsilon_{E}+\frac{\Delta_{X}}{H^{2}_{E}}\right)
\simeq-1+\frac{2}{3}\left(\epsilon_{E}-\frac{\Delta_{H}}{H_{E}}\right).
\end{equation}

Therefore, in the case of a weak coupling, a decrease in the energy density of a scalar field is negligible, the pressure and the state parameter changes only slightly on the inflationary stage of accelerated expansion of the universe.

The background dynamic equations (\ref{ex3})--(\ref{ex4}) in this case are reduced to expressions
\begin{eqnarray}
\label{BS1}
&&V_{GB}(\phi_{GB})= 3H^{2}_{E}+\dot{H}_{E}+H_{E}\Delta_{H}-2\Delta^{2}_{H}
\simeq3H^{2}_{E}+\dot{H}_{E}+H_{E}\Delta_{H},\\
\label{BS2}
&&\frac{1}{2}\dot{\phi}^{2}_{GB}=-\dot{H}_{E}-H_{E}\Delta_{H}-\Delta^{2}_{H}
\simeq-\dot{H}_{E}-H_{E}\Delta_{H}.
\end{eqnarray}

Also, on the basis of equation (\ref{connection}), one can write following expression
\begin{equation}
\label{BS3}
\dot{\xi}=\frac{H_{GB}-H_{E}}{2H^{2}_{GB}}=
\frac{1}{2H_{E}}\left[\frac{x}{(1+x)^{2}}\right]=
\frac{1}{2H_{E}}\left[x+{\mathcal O}(x^{2})+...\right]\simeq\frac{\Delta_{H}}{2H^{2}_{E}},
\end{equation}
where $x\equiv\frac{\Delta_{H}}{H_{E}}\ll1$.

Further, one can calculate the parameters of cosmological perturbations for the case of a weak GB coupling on the basis
of results which were obtained in Sec. \ref{GBPERT}.

The deviation parameters $\Delta_{1}$, $\Delta_{2}$ and $\Delta_{3}$ can be written from equation (\ref{DELTX})
and (\ref{DPCP1})--(\ref{DPCP3}) as
\begin{eqnarray}
\label{DPCP1WC}
&&\Delta_{1}=-2\frac{\Delta^{2}_{H}}{\Delta_{X}}\simeq2\frac{\Delta_{H}}{H_{E}},\\
\label{DPCP2WC}
&&\Delta_{2}=-2\frac{\Delta_{H}}{\Delta_{X}}\frac{d}{dt}\left(\frac{\Delta^{2}_{H}}{\Delta_{X}}\right)
\simeq-\frac{2}{H_{E}}\frac{d}{dt}\left(\frac{\Delta_{H}}{H_{E}}\right),\\
\label{DPCP3WC}
&&\Delta_{3}=2\frac{\Delta^{2}_{H}}{\Delta_{X}}-2\frac{d}{dt}\left(\frac{\Delta_{H}}{\Delta_{X}}\right)
\simeq-2\frac{\Delta_{H}}{H_{E}}-2\frac{\dot{H}_{E}}{H^{2}_{E}}=2\epsilon_{E}-2\frac{\Delta_{H}}{H_{E}}.
\end{eqnarray}

Further, after substituting (\ref{DPCP1WC}) and (\ref{DPCP3WC}) into (\ref{NSGB2}) one has
\begin{equation}
n_{S(GB)}-1=-2\epsilon_{E}-\frac{2}{H_{E}}\left(1+\frac{\Delta_{H}}{H_{E}}\right)^{-1}
\frac{d}{dt}\ln\left[\epsilon_{E}-\frac{\Delta_{H}}{H_{E}}\right],
\end{equation}
and, taking into account a weak GB coupling condition $\frac{\Delta_{H}}{H_{E}}\ll1$, the spectral index of a scalar perturbations can be written in following form
\begin{equation}
\label{NSGWC}
n_{S(GB)}-1\simeq-2\epsilon_{E}-\frac{2}{H_{E}}\frac{d}{dt}\ln\left[\epsilon_{E}-\frac{\Delta_{H}}{H_{E}}\right].
\end{equation}

Also, after substituting (\ref{DPCP1WC}) and (\ref{DPCP3WC}) into (\ref{RGB2}), one has expression for tensor-to-scalar ratio for a weak GB coupling
\begin{equation}
\label{RGBWC}
r_{GB}\simeq16\left[\epsilon_{E}-\frac{\Delta_{H}}{H_{E}}\right].
\end{equation}

As one can see, for the case $\Delta_{H}=0$, expressions (\ref{NSGWC})--(\ref{RGBWC}) are reduced to (\ref{NSE})--(\ref{RE}) for a standard inflation.

\subsection{Parametrization of a weak GB coupling influence on the inflationary process}

In general case, one can consider various types of the deviation parameter $\Delta_{H}$ to construct the models of cosmological inflation with non-minimal GB coupling. Nevertheless, it is possible to use the assumption that a weak GB-coupling does not change the type (or, otherwise, the shape) of the minimal coupled scalar field potential $V_{E}(\phi_{E})$, i.e. such a coupling does not change the nature of background inflationary processes occurring in early universe, which are determined by the type of the potential \cite{Martin:2013tda}.

In order to meet this assumption and compare such a models with ones based on Einstein gravity as well, one can define the deviation parameter as follows
\begin{equation}
\label{PARCON}
\Delta_{H}=-\alpha_{GB}\frac{\dot{H}_{E}}{H_{E}},
\end{equation}
where $\alpha_{GB}<1$ is a coupling constant.

For the case of quasi de Sitter expansion (when $\epsilon_{E}\ll1$), after substituting (\ref{PARCON}) into expressions for deviation parameters (\ref{DELTA1}) and (\ref{DELTA2}) and neglecting the small terms of second order and higher, from (\ref{WGVEL}) one has
\begin{equation}
\label{WGVELD}
c^{2}_{g}\simeq\frac{1+\alpha_{GB}\epsilon_{E}}{1-\alpha_{GB}\epsilon_{E}},~~~c_{g}\simeq1+\alpha_{GB}\epsilon_{E}.
\end{equation}

The state parameter of a scalar field (\ref{STATEPARAMETER}), for the deviation parameter (\ref{PARCON}), is
\begin{equation}
w_{GB}=-1+\frac{2}{3}\left(1-\alpha_{GB}\right)\epsilon_{E}.
\end{equation}

Thus, on inflationary stage, one has the deviations of the velocity of tensor perturbations from the speed of light in vacuum (in natural units) and deviations of the state parameter $w_{GB}$ from $w_{E}$, which are defined by the small factor $\alpha_{GB}\epsilon_{E}\ll1$ connected with deviations from pure de Sitter exponential expansion  (when $\epsilon_{E}=0$).

For pure exponential expansion from (\ref{PARCON}) one has $\Delta_{H}=0$ and $H_{GB}=H_{E}=const$, that implies  no differences between this type of a weak GB coupling and Einstein gravity, namely $w_{GB}=w_{E}=-1$ corresponding to Dark Energy, which can be defined by constant scalar field $\phi=const$ or cosmological constant $\Lambda$ associated with non-zero vacuum energy for the case $\phi=0$.
Consequently, for $\Lambda$CDM--model of a modern stage with accelerated exponential expansion of the universe \cite{Perlmutter:1998np,Riess:1998cb,Ade:2015xua} there are no deviations ($\Delta_{H}=0$) of this type of a weak GB coupling from Einstein gravity, which imply $c_{g}=1$ corresponding to restriction on the velocity of gravitational waves \cite{GBM:2017lvd}.

The background dynamic equations  (\ref{BS1})--(\ref{BS2}) with  the deviation parameter (\ref{PARCON}) can be written as
\begin{eqnarray}
\label{BS1WC}
&&V_{GB}(\phi_{GB})= 3H^{2}_{E}+(1-\alpha_{GB})\dot{H}_{E},\\
\label{BS2WC}
&&\dot{\phi}^{2}_{GB}=-2(1-\alpha_{GB})\dot{H}_{E}.
\end{eqnarray}

As one can see from equations (\ref{E1})--(\ref{E2}) and (\ref{BS1WC})--(\ref{BS2WC}), a weak GB-coupling defined by
the deviation parameter (\ref{PARCON}) does not change the shape of the potential.

From equations (\ref{WEAKC}), (\ref{PARCON}), (\ref{E1})--(\ref{E2}) and (\ref{BS1WC})--(\ref{BS2WC}) one has following connections between the potentials, Hubble parameters and scalar fields for standard inflation and EGB-inflation
\begin{eqnarray}
\label{CONV}
&&V_{GB}=V_{E}-\alpha_{GB}\dot{H}_{E},\\
\label{CONH}
&&H_{GB}= H_{E}(1+\alpha_{GB}\epsilon_{E}),\\
\label{CONHP}
&&\phi_{GB}=\phi_{E}\sqrt{1-\alpha_{GB}},
\end{eqnarray}
where a weak GB coupling (\ref{WEAKC}) implies that condition $\alpha_{GB}\epsilon_{E}\ll1$ is satisfied.

A coupling function can be obtained by integration of expression (\ref{BS3}) after substituting the deviation parameter (\ref{PARCON}) into this expression. As a result, one has
\begin{equation}
\label{CFGB}
\xi(\phi_{GB})=\frac{\alpha_{GB}}{4H^{2}_{E}(\phi_{GB})}+\xi_{0},
\end{equation}
where $\xi_{0}$ is the integration constant, however, it should be noted, that the constant coupling of a scalar field and the Gauss-Bonnet term $\xi=const$ does not affect the dynamic of four-dimensional Friedmann universe.

After neglecting the small terms $-\alpha_{GB}\dot{H}_{E}\ll V_{E}$ and $\alpha_{GB}\epsilon_{E}\ll1$ in expressions (\ref{CONV})--(\ref{CONH}) one has $V_{GB}\approx V_{E}$ and $H_{GB}\approx H_{E}$, that,
taking into account equations (\ref{BS1WC}) and (\ref{CFGB}), gives well known result \cite{Guo:2010jr,Koh:2014bka,vandeBruck:2017voa,Yi:2018gse} (with $\xi_{0}=0$)
\begin{equation}
\label{CFGBSR}
V_{GB}(\phi_{GB})[\xi(\phi_{GB})-\xi_{0}]\approx\frac{3}{4}\alpha_{GB}=const,
\end{equation}
therefore, in this approximation one has the changing of a field (\ref{CONHP}) only.

To determine the difference between results (\ref{CFGB}) and (\ref{CFGBSR}), one can consider a nominal coupling function in the following form
\begin{equation}
\label{BS5}
\xi(\phi_{GB})\simeq\frac{3}{4}\alpha_{GB}\left[V_{GB}(\phi_{GB})-(1-\alpha_{GB})\dot{H}_{E}
+\Lambda\right]^{-1}+\xi_{0}=\frac{3}{4}\alpha_{GB}\left[V_{GB}(\phi_{GB})+X_{GB}+\Lambda\right]^{-1}+\xi_{0},
\end{equation}
where potential was redefined as
\begin{equation}
\label{POTRED}
V_{GB}(\phi_{GB})\rightarrow V_{GB}(\phi_{GB})+\Lambda,
\end{equation}
and cosmological constant $\Lambda$ can be associated with non-zero vacuum energy.

From (\ref{CFGBSR}) one has expression similar to (\ref{BS5}) with $X_{GB}=0$ after redefinition (\ref{POTRED}), which was considered in~\cite{vandeBruck:2017voa,Yi:2018gse} to eliminate the divergence in a value of non-minimal coupling function $\xi_{GB}(\phi_{GB})$ after completion of inflationary stage.

From expression (\ref{BS5}) on the inflationary stage when $V_{GB}\gg X_{GB}+\Lambda$ one has
\begin{equation}
\label{BS7}
\xi(\phi_{GB})\simeq\frac{3\alpha_{GB}}{4V_{GB}(\phi_{GB})}+\xi_{0},
\end{equation}
and at the reheating phase when $X_{GB}\gg V_{GB}+\Lambda$ and energy of a scalar field is transferred to radiation, the coupling function is
\begin{equation}
\label{BS8}
\xi(\phi_{GB})\simeq\frac{3\alpha_{GB}}{4X_{GB}}+\xi_{0}.
\end{equation}

At the following stage, when $\Lambda\gg X_{GB}+V_{GB}$ one has
\begin{equation}
\label{BS9}
\xi(\phi_{GB})\simeq\frac{3\alpha_{GB}}{4\Lambda}+\xi_{0}=const,
\end{equation}
that implies the negligible GB coupling effects after completion of inflationary stage.

Therefore, the difference with results obtained in \cite{Guo:2010jr,Koh:2014bka,vandeBruck:2017voa,Yi:2018gse} is the existence of an additional stage of the predominance of  a scalar field's kinetic energy (kination) with corresponding coupling function $\xi(\phi_{GB})$ which is defined by expression (\ref{BS8}).
It should be noted that taking into account the stage of the predominance of kinetic energy leads to additional dynamic effects beyond the slow-roll regime, which will be shown further for specific EGB-inflationary models.

\subsection{The dynamic equations in terms of a scalar field}\label{DESF}

To generate the exact solutions for the inflationary models based on Einstein gravity one can use the equations (\ref{E1})--(\ref{E2}), and, moreover, the method based on the representation of dynamic equations in terms of a scalar field
\footnote{A review of this method is given in \cite{Chervon:2017kgn,Chervon:2019sey}.}.

In the framework of this approach, one can rewrite the background dynamic equations for standard inflation (\ref{E1})--(\ref{E2}) on the basis of the relation
\begin{equation}
\dot{H}_{E}=\frac{dH_{E}}{dt}=\frac{dH_{E}}{d\phi_{E}}\frac{d\phi_{E}}{dt}=
\frac{dH_{E}}{d\phi_{E}}\dot{\phi}_{E},
\end{equation}
in following form
\begin{eqnarray}
\label{ISB1}
&&V_{E}(\phi_{E})= 3H^{2}_{E}-2\left(\frac{dH_{E}}{d\phi_{E}}\right)^{2},\\
\label{ISB2}
&&\dot{\phi}_{E}=-2\frac{dH_{E}}{d\phi_{E}}.
\end{eqnarray}

Also, the slow-roll parameters can be determined from expressions (\ref{EPSDELE}) as
\begin{eqnarray}
\label{EPSPIHI}
&&\epsilon_{E}=-\frac{\dot{H}_{E}}{H^{2}_{E}}=2H^{-2}_{E}\left(\frac{dH_{E}}{d\phi_{E}}\right)^{2},\\
\label{DELTAPHI}
&&\delta_{E}=-\frac{\ddot{H}_{E}}{2\dot{H}_{E}H_{E}}=2H^{-1}_{E}\left(\frac{dH^{2}_{E}}{d\phi_{E}^{2}}\right).
\end{eqnarray}

In this case, one can generate the exact solutions of equations (\ref{ISB1})--(\ref{ISB2}) by the choice of the Hubble parameter
$H_{E}=H_{E}(\phi_{E})$ as the function of a scalar field $\phi_{E}$.

For the case of a weak GB coupling, on the basis of relation
\begin{equation}
\dot{H}_{E}=\frac{dH_{E}}{d\phi_{GB}}\dot{\phi}_{GB},
\end{equation}
from equations (\ref{BS1WC})--(\ref{BS2WC}) one has
\begin{eqnarray}
\label{ISB1GB}
&&V_{GB}(\phi_{GB})= 3H^{2}_{E}-2(1-\alpha_{GB})^{2}\left(\frac{dH_{E}}{d\phi_{GB}}\right)^{2},\\
\label{ISB2GB}
&&\dot{\phi}_{GB}=-2(1-\alpha_{GB})\frac{dH_{E}}{d\phi_{GB}},
\end{eqnarray}
where the Hubble parameter $H_{E}=H_{E}(\phi_{GB})$ is considered as the function of a scalar field $\phi_{GB}$.

As one can see from equations (\ref{ISB1})--(\ref{ISB2}) and (\ref{ISB1GB})--(\ref{ISB2GB}), a weak GB coupling for $f_{GB}(t)=\alpha_{GB}=const$ doesn't change the shape of the potential $V_{GB}$ compared with $V_{E}$ for inflation based on Einstein gravity, which is a specific property of such a choice of the function $f_{GB}$ or the deviation parameter $\Delta_{H}$.

The expression for the non-minimal coupling function remains the same
\begin{equation}
\label{CFGBISB}
\xi(\phi_{GB})=\frac{\alpha_{GB}}{4H^{2}_{E}(\phi_{GB})}+\xi_{0},
\end{equation}
and the connection (\ref{CONH}) between Hubble parameters $H_{GB}$ and $H_{E}$ can be written as
\begin{equation}
\label{CONHPHI}
H_{GB}=H_{E}(1+\alpha_{GB}\epsilon_{E})=
H_{E}\left[1+2\alpha_{GB}H^{-2}_{E}\left(\frac{dH_{E}}{d\phi_{E}}\right)^{2}\,\right],
\end{equation}
with the same relation (\ref{CONHP}) between scalar fields $\phi_{GB}$ and $\phi_{E}$.

Thus, expressions (\ref{BS1WC})--(\ref{CFGB}) or (\ref{ISB1})--(\ref{CONHPHI}) completely determine the relations between exact inflationary solutions in the case of Einstein gravity (see, for example, in  \cite{Fomin:2017xlx,Chervon:2017kgn,Fomin:2018uql,Chervon:2019sey}) and approximate ones for a weak GB coupling. The difference between the solutions is determined by a coupling constant $\alpha_{GB}$.

\subsection{The parameters of cosmological perturbations for a weak GB coupling}

After substituting (\ref{PARCON}) into equations (\ref{NSGWC})--(\ref{RGBWC}) one has following expressions for the parameters of cosmological perturbations corresponding to the deviation parameter (\ref{PARCON}) for a weak GB coupling
\begin{eqnarray}
\label{NSGWC2}
&&n_{S(GB)}-1=-4\epsilon_{E}+2\delta_{E}=n_{S(E)},\\
\label{RGBWC2}
&&r_{GB}=16(1-\alpha_{GB})\epsilon_{E}=(1-\alpha_{GB})r_{E}.
\end{eqnarray}

It should be noted that a similar expressions of the parameters of cosmological perturbations for EGB-inflation
\begin{eqnarray}
\label{NSGWC2SR}
&&n_{S(GB)}-1=-2\epsilon_{1}-\epsilon_{2},\\
\label{RGBWC2SR}
&&r_{GB}=16(1-\lambda)\epsilon_{1},
\end{eqnarray}
were considered earlier in the paper \cite{Yi:2018gse} on the basis of postulated connection $\delta_{1}=2\lambda\epsilon_{1}$ between slow-roll parameters, where $\lambda$ is a some constant, $\delta_{1}=\Delta_{1}$, $\epsilon_{1}=\epsilon_{GB}$ and $\epsilon_{2}=2(\epsilon_{GB}-\delta_{GB})$.

The difference between expressions (\ref{NSGWC2})--(\ref{RGBWC2}) and (\ref{NSGWC2SR})--(\ref{RGBWC2SR}) is that  the first ones (\ref{NSGWC2})--(\ref{RGBWC2}) determine the difference between cosmological perturbations parameters for inflationary models based on General Relativity and Einstein-Gauss-Bonnet gravity, the second expressions (\ref{NSGWC2SR})--(\ref{RGBWC2SR}) correspond to the case of Einstein-Gauss-Bonnet gravity only where the difference between EGB-inflationary models is defined by the value of the parameter $\lambda$.

The transition from expressions (\ref{NSGWC2SR})--(\ref{RGBWC2SR}) to (\ref{NSGWC2})--(\ref{RGBWC2}) can be carried out as follows: after neglecting the small term $\alpha_{GB}\epsilon_{E}\ll1$ in equation (\ref{CONH}) one has $H_{GB}\approx H_{E}$, which implies the following relations $\epsilon_{GB}\approx\epsilon_{E}$ and $\delta_{GB}\approx\delta_{E}$. Further, after substituting these relations into (\ref{NSGWC2SR})--(\ref{RGBWC2SR}), one has (\ref{NSGWC2})--(\ref{RGBWC2}).
Also, for the case $\alpha_{GB}=0$ and $\lambda=0$, all these expressions are reduced to (\ref{NSE})--(\ref{RE})
corresponding to Einstein gravity.

As one can see, for a weak GB coupling, the corrections to the value of the spectral index of scalar perturbations are
negligible $n_{S(GB)}\simeq n_{S(E)}$. Nevertheless, such a coupling can have a significant effect on the value of the tensor-to-scalar ratio, namely, the positive coupling constant $0<\alpha_{GB}<1$ leads to decreasing the value
of tensor-to-scalar ratio $r_{GB}<r_{E}$, and the negative one $\alpha_{GB}<0$ gives
a greater contribution of tensor perturbations to the CMB anisotropy than in the case of standard inflation $r_{GB}>r_{E}$.

Thus, the corrections to Einstein gravity associated with a weak non-minimal coupling of a scalar field and the Gauss-Bonnet term can have a significant effect on the verification of cosmological models from observational constraints.

\section{The examples of inflationary models with a weak GB coupling}\label{section5}

In order to determine in more detail the influence of a weak GB coupling on the inflationary parameters, one can consider known models of cosmological inflation based on Einstein gravity with corrections (\ref{CONV})--(\ref{CONHP}).
A description of the inflationary models under consideration can be found, for example, in the reviews \cite{Martin:2013tda,Chervon:2017kgn,Chervon:2019sey}, and in many other papers as well.
It should be noted that transformation (\ref{POTRED}) can be applied to all models under consideration to eliminate the divergence in a value of non-minimal coupling function $\xi(\phi_{GB})$ after completion of inflationary stage.

Also, one can use the observational constraints on the values of cosmological perturbation parameters (\ref{PLANCK1})--(\ref{PLANCK2}) to estimate the value of a coupling constant $\alpha_{GB}$.
At the level of qualitative analysis, from expressions (\ref{NSGWC2})--(\ref{RGBWC2}), one has that
for the case  $\alpha_{GB}\leq0$, the standard inflationary models and corresponding EGB-inflation can be verified,
and for $0<\alpha_{GB}<1$,  EGB-inflation only corresponds to the observational constraints.
On the other hand, quantitative estimates of the coupling constant $\alpha_{GB}$ make it possible to determine the influence of a weak GB coupling on the parameters of cosmological inflationary models.

\subsection{Power-law inflation with a weak GB coupling}

For power-law inflation based on Einstein gravity \cite{Martin:2013tda,Chervon:2017kgn,Chervon:2019sey} the Hubble parameter and corresponding scale factor are
\begin{eqnarray}
\label{HCIPL}
&&H_{E}(t)=\frac{n}{t},\\
\label{ACIPL}
&&a_{E}(t)=a_{0}t^{n},
\end{eqnarray}
where $n$ is the positive constant and $a_{0}$ is initial value of a scale factor.

The Hubble parameter and scale factor for EGB-inflation one can obtain from equation (\ref{CONH}) as
\begin{eqnarray}
\label{GBPL}
&&H_{GB}(t)=\frac{n+\alpha_{GB}}{t},\\
\label{GBPL1}
&&a_{GB}(t)=a_{0}t^{(n+\alpha_{GB})}.
\end{eqnarray}

From equations (\ref{BS1WC})--(\ref{BS2WC}) one has the following expressions for the scalar field and potential
\begin{eqnarray}
\label{HPLGB}
&&\phi_{GB}(t)=\pm\sqrt{2n(1-\alpha_{GB})}\ln t,\\
\label{VGBPL}
&&V_{GB}(\phi_{GB})=n(3n-1+\alpha_{GB})\exp\left[\mp\sqrt{\frac{2}{n(1-\alpha_{GB})}}\,\phi_{GB}\right],
\end{eqnarray}
taking into account transformation (\ref{POTRED}).

The coupling function (\ref{CFGB}) for the power-law EGB-inflation is
\begin{equation}
\label{CFPL}
\xi_{GB}(\phi_{GB})=\left(\frac{\alpha_{GB}}{n}\right)\exp\left[\pm\sqrt{\frac{2}{n(1-\alpha_{GB})}}\,\phi_{GB}\right]
+\xi_{0}.
\end{equation}

For the case $\alpha_{GB}=0$ one has $\xi=0$, and solutions (\ref{HPLGB})--(\ref{VGBPL}) are transformed to standard inflationary ones $\phi_{GB}\rightarrow\phi_{E}$ and $V_{GB}\rightarrow V_{E}$ corresponding to (\ref{HCIPL})--(\ref{ACIPL}), namely
\begin{eqnarray}
\label{HPLE}
&&\phi_{E}(t)=\pm\sqrt{2n}\ln t,\\
\label{VEPL}
&&V_{E}(\phi_{E})=n(3n-1)\exp\left[\mp\sqrt{\frac{2}{n}}\,\phi_{E}\right].
\end{eqnarray}

For the Hubble parameter (\ref{HCIPL}) from expressions (\ref{EPSDELE}) one has $\epsilon_{E}=\delta_{E}$, therefore, the connection between tensor-to-scalar ratio and spectral index of scalar perturbations can be noted on the basis of (\ref{NSGWC2})--(\ref{RGBWC2}) as
\begin{equation}
\label{NSRPL}
r=8(1-\alpha_{GB})(1-n_{S}).
\end{equation}

From  observational constraints (\ref{PLANCK1})--(\ref{PLANCK2}) one can obtain that for the verifiable power-law inflation
the values of a coupling constant are in the range $0.8<\alpha_{GB}<1$. Therefore, from expression (\ref{CONHP}) one has the following estimations for the non-minimal coupled scalar field $0<|\phi_{GB}|<0.45|\phi_{E}|$. Since the coupling constant can take the positive values $\alpha_{GB}>0$ only, this type of cosmological inflation is not verified for the case of Einstein gravity.

\subsection{The quadratic chaotic inflation with a weak GB coupling}\label{QCIS}

For quadratic chaotic inflation with a massive scalar field \cite{Linde:1983gd,Martin:2013tda,Chervon:2017kgn,Chervon:2019sey} the Hubble parameter can be defined as
\begin{equation}
\label{QCH}
H_{E}(t)=-\frac{m^{2}_{E}}{3}t+\frac{m_{E}\phi_{0}}{\sqrt{6}},
\end{equation}
where $m_{E}$ is the mass of a scalar field.

From equations (\ref{E1})--(\ref{E2}) one has following scalar field and it's potential
\begin{eqnarray}
\label{QCPH}
&&\phi_{E}(t)=-m_{E}\sqrt{\frac{2}{3}}t+\phi_{0},\\
\label{QCV}
&&V_{E}(\phi_{E})=\frac{m^{2}_{E}\phi^{2}_{E}}{2}-\frac{m^{2}_{E}}{3},
\end{eqnarray}
where $\phi_{E}$ and $\phi_{0}$ are the scalar field and it's initial value.

For the Hubble parameter (\ref{QCH}) one has the following corresponding scale factor
\begin{equation}
a_{E}(t)=a_{0}\exp\left[\frac{m_{E}t}{6}\left(\sqrt{6}\phi_{0}-m_{E}t\right)\right].
\end{equation}

For EGB-inflation, from equations (\ref{BS1WC})--(\ref{BS2WC}),  one has
\begin{eqnarray}
\label{QCPHGB}
&&\phi_{GB}(t)=\sqrt{1-\alpha_{GB}}\left(-m_{E}\sqrt{\frac{2}{3}}t+\phi_{0}\right),\\
\label{QCV2}
&&V_{GB}(\phi_{GB})=\frac{m^{2}_{GB}\phi^{2}_{GB}}{2}-\frac{m^{2}_{GB}}{3}(1-\alpha_{GB})^{2},\\
\label{QCXI}
&&\xi(\phi_{GB})=\frac{3\alpha_{GB}}{2m^{2}_{GB}}\phi^{-2}_{GB},
\end{eqnarray}
where the masses of a scalar field for the case of Einstein-Gauss-Bonnet gravity and General Relativity are related as
follows
\begin{equation}
\label{MASSES}
m_{GB}=\frac{m_{E}}{\sqrt{1-\alpha_{GB}}},
\end{equation}
since the deviation parameter
\begin{equation}
\Delta_{V}=V_{GB}-V_{E}=-\alpha_{GB}\dot{H}_{E}=\alpha_{GB}\frac{m^{2}_{E}}{3}=const,
\end{equation}
and, therefore, from (\ref{QCV}) and (\ref{QCV2}) one has the condition $m_{E}\phi_{E}=m_{GB}\phi_{GB}$, that,
 taking into account (\ref{CONHP}), leads to relation (\ref{MASSES}).

The Hubble parameter for EGB-inflation one can obtain from (\ref{CONH}) and (\ref{QCH}) as
\begin{equation}
\label{QCHGB}
H_{GB}(t)=-\frac{m^{2}_{E}}{3}t+\frac{m_{E}\phi_{0}}{\sqrt{6}}+
\frac{2\alpha_{GB}m_{E}}{\sqrt{6}\phi_{0}-2m_{E}t},
\end{equation}
with corresponding scale factor
\begin{equation}
\label{aGB}
a_{GB}(t)=a_{0}\left(\sqrt{6}\phi_{0}-2m_{E}t\right)^{-\alpha_{GB}}
\exp\left[\frac{m_{E}t}{6}\left(\sqrt{6}\phi_{0}-m_{E}t\right)\right],
\end{equation}
therefore, the cosmic time of inflationary stage is restricted by following value $t_{inf}<\sqrt{\frac{2}{3}}\frac{\phi_{0}}{m_{E}}$.

Thus, for this type of inflation, the non-minimal coupling of a scalar field and the Gauss-Bonnet scalar changes the mass of the field. The mass of the scalar field $\phi_{GB}$ depends on the value of a coupling parameter $\alpha_{GB}$, which can be estimated by means of the observational constraints (\ref{PLANCK1})--(\ref{PLANCK2}) on the values of cosmological perturbation parameters (\ref{NSGWC2})--(\ref{RGBWC2}).

 From (\ref{EPSDELE}) it follows that the second slow-roll parameter corresponding to the Hubble parameter (\ref{QCH}) is $\delta_{E}=0$, and, therefore, from equations (\ref{NSGWC2})--(\ref{RGBWC2}) one has the connection between tensor-to-scalar ratio and spectral index of scalar perturbations, which can be noted as
\begin{equation}
\label{QCRNS}
r=4(1-\alpha_{GB})(1-n_{S}).
\end{equation}

From expressions (\ref{PLANCK1})--(\ref{PLANCK2}) and relation (\ref{QCRNS}) one can conclude that the model of chaotic inflation with quadratic potential corresponds to the observational constraints for $0.45<\alpha_{GB}<1$. Therefore, the mass of the scalar field non-minimally coupled with the Gauss-Bonnet scalar for verifiable quadratic chaotic inflation can be estimated as $m_{GB}>1.35\,m_{E}$, i.e. the mass of the field increases. From expression (\ref{CONHP}) one has that the field itself changes as $0<\phi_{GB}<0.74\,\phi_{E}$.

\subsection{The Higgs inflation with a weak GB coupling}

For this type of inflation \cite{Linde:1983gd,Lyth:2007qh,Mazumdar:2010sa,Yamaguchi:2011kg}, one can consider the Hubble parameter
\begin{equation}
\label{HHIGGS}
H_{E}(t)=\beta\left[\frac{2}{3}\mu+\phi^{2}_{0}\exp(-8\beta t)\right],
\end{equation}
with corresponding scale factor
\begin{equation}
\label{AHIGGS}
a_{E}(t)=a_{0}\exp\left(\frac{2\mu}{3}\beta t+\frac{\phi^{2}_{0}}{8}e^{-8\beta t}\right),
\end{equation}
where $\beta$ and $\mu$ are some constants.

From equations (\ref{E1})--(\ref{E2}) one has
\begin{eqnarray}
\label{PHHIGGS}
&&\phi_{E}(t)=\phi_{0}\exp(-4\beta t),\\
\label{VHIGGS}
&&V_{E}(\phi_{E})=\frac{\lambda_{E}}{9}\mu^{2}+\frac{1}{2}m^{2}_{E}\phi^{2}_{E}+
\frac{\lambda_{E}}{4}\phi^{4}_{E},
\end{eqnarray}
where $\phi_{0}$ is the initial value of the scalar field, $\lambda_{E}=12\beta^{2}$ is the self-coupling constant, and squared mass of the field is $m^{2}_{E}=\frac{2}{3}\lambda_{E}(\mu-2)$.

For the case $\mu<2$ one has $m^{2}_{E}<0$ that corresponds to the spontaneously broken symmetry in this model \cite{Linde:1983gd,Lyth:2007qh,Mazumdar:2010sa,Yamaguchi:2011kg}.
For the case $\mu>2$ one has a model without the symmetry breaking where squared mass is $m^{2}_{E}>0$, and the value $\mu=2$ implies the transition from the Higgs inflation to chaotic one with potential $V_{E}\sim\phi^{4}_{E}$ \cite{Linde:1983gd}.

For EGB-inflation, from equations (\ref{BS1WC})--(\ref{BS2WC}), one has
\begin{eqnarray}
\label{HIGGSPGB}
&&\phi_{GB}(t)=\phi_{0}\sqrt{1-\alpha_{GB}}\exp(-4\beta t),\\
\label{HIGGSVGB}
&&V_{GB}(\phi_{GB})=\frac{\lambda_{GB}}{9}(1-\alpha_{GB})^{2}\mu^{2}+
\frac{1}{2}m^{2}_{GB}\phi^{2}_{GB}+\frac{\lambda_{GB}}{4}\phi^{4}_{GB},\\
\label{HIGGSXIGB}
&&\xi_{GB}(\phi_{GB})=27\left(\frac{\alpha_{GB}}{\lambda_{GB}}\right)
\left[3\phi^{2}_{GB}-2\mu(1-\alpha_{GB})\right]^{-2},
\end{eqnarray}
where
\begin{eqnarray}
\label{LAMBDAGB}
&&\lambda_{GB}=\frac{12\beta^{2}}{(1-\alpha_{GB})^{2}}=\frac{\lambda_{E}}{(1-\alpha_{GB})^{2}},\\
\label{SIGMAGB}
&&m^{2}_{GB}=\frac{2}{3}\lambda_{E}\left(\frac{\mu+2\alpha_{GB}-2}{1-\alpha_{GB}}\right).
\end{eqnarray}


Also, the mass of the Higgs field coupled with Gauss-Bonnet term can be defined as
 \begin{equation}
\label{MASSHIGGS}
m^{2}_{GB}=m^{2}_{E}+\frac{2}{3}\mu\lambda_{E}\left(\frac{\alpha_{GB}}{1-\alpha_{GB}}\right).
\end{equation}

The condition of spontaneous symmetry breaking also change, namely, taking into account a weak GB coupling, such a condition can be written as
 \begin{equation}
\label{SBS}
\mu+2\alpha_{GB}-2<0.
\end{equation}

Also, for the case $\mu+2\alpha_{GB}-2=0$ one has the transition from the Higgs inflation to chaotic EGB-inflation with corresponding potential $V_{GB}\sim\phi^{4}_{GB}$.

The Hubble parameter and the scale factor for the Higgs inflation with a weak GB coupling are
\begin{eqnarray}
\label{HHIGGSGB}
&&H_{GB}(t)=H_{E}(t)+\frac{24\alpha_{GB}\beta\phi^{2}_{0}e^{-8\beta t}}{2\mu+3\phi^{2}_{0}e^{-8\beta t}},\\
\label{AHIGGSGB}
&&a_{GB}(t)= a_{E}(t)K_{GB}e^{8\alpha_{GB}\beta t}\left(3\phi^{2}_{0}+2\mu e^{8\beta t}\right)^{-\alpha_{GB}},
\end{eqnarray}
where $K_{GB}$ is the constant of integration.

As one can see, for $2\mu e^{8\beta t}\gg 3\phi^{2}_{0}$ one has $a_{GB}(t)\simeq a_{E}(t)$ up to constant $2^{-\alpha_{GB}}K_{GB}$, that corresponds the condition $\alpha_{GB}\epsilon_{E}\ll1$ in expression (\ref{CONH}) for $2^{-\alpha_{GB}}K_{GB}\simeq1$.

For the Hubble parameter (\ref{HHIGGS}) from expressions (\ref{EPSDELE})  one has
$\epsilon_{E}=-\frac{1}{3}\delta^{2}_{E}+2\delta_{E}\simeq2\delta_{E}$, and the connection between tensor-to-scalar ratio and spectral index of scalar perturbations for the Higgs inflation is
\begin{equation}
\label{NSRHIGGS}
r=\frac{16}{3}(1-\alpha_{GB})(1-n_{S}).
\end{equation}

From  observational constraints (\ref{PLANCK1})--(\ref{PLANCK2}) one has the following values of a coupling constant $0.6<\alpha_{GB}<1$ for the verifiable inflation with the Higgs potential. Thus, the mass of the Higgs field for a weak GB coupling changes as $m^{2}_{GB}>m^{2}_{E}+\mu\lambda_{E}$ and self-coupling parameter is $\lambda_{GB}>1.58\lambda_{E}$. The scalar field for this type of inflation with a weak GB coupling can be estimated as $0<\phi_{GB}<0.63\,\phi_{E}$.

Another possibility of constructing verified Higgs inflation is to consider the non-minimal coupling of a scalar field with the Ricci scalar, which, however, changes the shape of the Higgs potential \cite{Bezrukov:2007ep,Mishra:2018dtg} in contrast to a weak GB coupling.

\subsection{Intermediate inflation with a weak GB coupling}

For the approach presented in Sec. \ref{DESF}, the intermediate inflation \cite{Martin:2013tda,Chervon:2017kgn,Chervon:2019sey,Muslimov:1990be} can be considered on the basis of the Hubble parameter
\begin{equation}
\label{INTHPH}
H_{E}(\phi_{E})=A\phi^{-k}_{E},
\end{equation}
where $A$ and $k$ are some positive constants.

From equations (\ref{ISB1})--(\ref{ISB2}) for the Hubble parameter (\ref{INTHPH}) one has following cosmological solutions
for the case of Einstein gravity
\begin{eqnarray}
\label{INTIV}
&&V_{E}(\phi_{E})=3A^{2}\phi^{-2k}_{E}-2A^{2}k^{2}\phi^{-2(k+1)}_{E},\\
\label{INTIPHI}
&&\phi_{E}(t)=\left[c+2Ak(k+2)t\right]^{\frac{1}{k+2}},\\
\label{INTIHT}
&&H_{E}(t)=A\left[c+2Ak(k+2)t\right]^{-\frac{k}{k+2}},\\
\label{INTIA}
&&a_{E}(t)=a_{0}\exp\left(\frac{1}{4k}\left[c+2Ak(k+2)t\right]^{\frac{2}{k+2}}\right),
\end{eqnarray}
where $c$ is the constant of integration.

As one can see, the expansion of the universe, in this case, is faster than a power-law and slower than a pure exponential one.

For EGB-inflation with the Hubble parameter
\begin{equation}
\label{INTHGB}
H_{E}(\phi_{GB})=\tilde{A}\phi^{-k}_{GB},
\end{equation}
where $\tilde{A}=A(1-\alpha_{GB})^{k/2}$, from equations (\ref{ISB1GB})--(\ref{CFGBISB}) one has
\begin{eqnarray}
\label{INTIVGB}
&&V_{GB}(\phi_{GB})=3\tilde{A}^{2}\phi^{-2k}_{GB}-2\tilde{A}^{2}k^{2}
(1-\alpha_{GB})^{2}\phi^{-2(k+1)}_{GB},\\
\label{INTIPHIGB}
&&\phi_{E}(t)=\sqrt{1-\alpha_{GB}}\left[c+2Ak(k+2)t\right]^{\frac{1}{k+2}},\\
\label{INTIPXIIGB}
&&\xi(\phi_{GB})=\frac{\alpha_{GB}}{4\tilde{A}^{2}}\phi^{2k}_{GB}.
\end{eqnarray}

The Hubble parameter for EGB-inflation can be defined from (\ref{CONHPHI}) and (\ref{INTIPHIGB}) as follows
\begin{equation}
\label{INTHPHGB}
H_{GB}(t)=H_{E}(t)+\frac{2Ak^{2}\alpha_{GB}}{c+2Ak(k+2)t},
\end{equation}
with corresponding scale factor
\begin{equation}
\label{INTIAGB}
a_{GB}(t)=a_{E}(t)\tilde{c}\left[c+2Ak(k+2)t\right]^{\alpha_{GB}\left(\frac{k}{k+2}\right)},
\end{equation}
where $\tilde{c}$ is the constant of integration.

Thus, a non-minimal GB coupling leads to the appearance of an additional power-law term in the scale factor, which corresponds to the possibility of the exit from the inflationary stage of accelerated expansion of the universe.

After substituting the Hubble parameter (\ref{INTHPH}) into expressions (\ref{EPSPIHI})--(\ref{DELTAPHI}) one has the following relation between slow-roll parameters
\begin{equation}
\label{SRCONINTI}
\delta_{E}=\left(\frac{k+1}{k}\right)\epsilon_{E}.
\end{equation}

Thus, the connection between tensor-to-scalar ratio and spectral index of scalar perturbations for this type of inflation
on the basis of equations  (\ref{NSGWC2})--(\ref{RGBWC2}) and (\ref{SRCONINTI}) can be written as
\begin{equation}
\label{RNSINTI}
r=8k\left(\frac{1-\alpha_{GB}}{k-1}\right)(1-n_{S}).
\end{equation}

In this case, the tensor-to-scalar ratio is defined by two parameters $\alpha_{GB}$ and $k$, and the relation between these parameters is determined by the following inequality
\begin{equation}
\label{RNSINCONGB}
0<k\left(\frac{1-\alpha_{GB}}{k-1}\right)<\frac{0.065}{8(1-n_{S})}<0.215.
\end{equation}

Inequality (\ref{RNSINCONGB}) shows that standard intermediate inflation ($\alpha_{GB}=0$) with $k>0$ doesn't
correspond to the observational constraints (\ref{PLANCK1})--(\ref{PLANCK2}). Thus, the coupling constant $\alpha_{GB}$ can have the positive values only. Also, conditions $0<\alpha_{GB}<1$ and (\ref{RNSINCONGB}) correspond to the values $k>1$.

For example, for $k=2$ from (\ref{RNSINCONGB}) one has the restrictions on the values of a coupling constant $0.9<\alpha_{GB}<1$, that leads to $0<\phi_{GB}<0.3\,\phi_{E}$.
Similarly, one can find constraints on a coupling constant for other values of the parameter $k$
in the case of intermediate inflation.

Also, it is possible to estimate the influence of a weak GB coupling for the other standard models of cosmological inflation based on Einstein gravity (see, for example, in \cite{Martin:2013tda,Chervon:2017kgn,Gron:2018rtj,Chervon:2019sey}) by using this approach.

\section{Discussion}\label{discussion}

In this paper, the influence of the non-minimal coupling of a scalar field and the Gauss-Bonnet scalar on the process of cosmological inflation was considered. The basis of the proposed analysis is the presented relations between the parameters of inflationary models for the case of General Relativity and Einstein-Gauss-Bonnet gravity.

The effect of non-minimal GB coupling was determined by means of  background deviation parameters $\{\Delta_{H},\Delta_{X},\Delta_{V}\}$ and ones corresponding to the Gauss-Bonnet term corrections to the parameters of cosmological perturbations $\{\Delta_{1},\Delta_{2},\Delta_{3}\}$ which are related by equations (\ref{REL}) and (\ref{DPCP1})--(\ref{DPCP3}).

In general case, two main results were obtained that characterize the influence of the GB-term on the inflationary process.
The first result is that in four-dimensional spatially flat Friedmann-Robertson-Walker space-time the non-minimal coupling of a scalar field with the Gauss-Bonnet term leads to decrease of it's energy density.
The second result is that such a coupling can accelerate the rate of expansion of the universe by less than two times.

Nevertheless, the slow-roll conditions correspond to a weak influence of such a coupling on cosmological dynamics.
This result leads to the interpretation of non-minimal coupling of a scalar field and the Gauss-Bonnet scalar as a small quantum corrections to the main dynamical effects of General Relativity at the inflationary stage of the evolution of early universe.

Based on this notion, which was called a weak GB coupling, the effect of such a coupling on the background inflationary parameters and cosmological perturbation parameters was evaluated. The influence of a weak GB coupling was also parameterized by means of a special choice of the deviation parameter $\Delta_{H}$, implying the conservation of the shape of minimal coupled scalar field potential $V_{E}$.
The result of this parameterization was that the influence of the Gauss-Bonnet scalar on the inflationary parameters is determined by the value of the coupling constant $\alpha_{GB}$ only.
It should also be noted that in the case of exponentially accelerated expansion of the universe, the difference between EGB gravity and General relativity is absent for any value of the parameter $\alpha_{GB}$ for a weak GB coupling under consideration.

The first effect of such a coupling is a change of a scalar field itself, which decreases for positive values of a coupling constant $\alpha_{GB}$. The second effect is a change in the mass of a scalar field in the inflationary model with quadratic potential and for the Higgs inflation based on Einstein-Gauss-Bonnet gravity compared to General Relativity.  Also, the non-minimal coupling of the Higgs field with the Gauss-Bonnet term changes the condition of spontaneous symmetry breaking for the Higgs inflation. The third effect is arising the additional terms in a scale factor which have a weak influence at the inflationary stage, however, they can change the cosmological dynamics beyond the slow-roll regime.
The observational constraints on the parameters of cosmological perturbations were used to estimate the value of a coupling constant $\alpha_{GB}$ for different inflationary models.

Thus, the proposed approach, on the one hand, allows to define the Gauss-Bonnet term corrections for any standard inflationary model based on Einstein gravity; on the other hand, it gives the possibility to verify such a models from observational constraints on the parameters of cosmological perturbations.

\begin{acknowledgments}
The study was funded by RFBR grant 20-02-00280 A.
\end{acknowledgments}

\bibliography{ref_GB}

\end{document}